\documentclass[english,prb,floatfix,letterpaper]{revtex4-2}

\usepackage[T1]{fontenc}
\usepackage[latin9]{inputenc}
\usepackage{amsmath}
\usepackage{amssymb}
\usepackage{babel}
\usepackage{url}

\usepackage{graphics}
\usepackage{graphicx}
\usepackage{epsfig}

\begin{document}

\title{Exploring phases of the Su-Schrieffer-Heeger model with tSNE}

\author{R. M. Woloshyn}
\affiliation{TRIUMF, 4004 Wesbrook Mall, Vancouver,
British Columbia, Canada V6T 2A3}

\begin{abstract}
T-distributed stochastic neighborhood embedding (tSNE) is used as a tool
to reveal the phase diagram of the Su-Schrieffer-Heeger model and some of
its extended and non-Hermitian variants. 
Bloch vectors calculated at different points in the
parameter space are mapped to a two-dimensional reduced space. The clusters
in the reduced space are used to visualize different phase regions included
in the input. The tSNE mapping is shown to be effective even in the challenging
case of the non-Hermitian extended model where five different phases are 
present. An example of using wavefunction input, instead of Bloch vectors,
is presented also.

\end{abstract}

\maketitle

\section{INTRODUCTION}
\label{intro}

Machine learning is being increasingly utilized in physics \cite{Mehta_2019,Carleo_2019}. 
Examples of applications include event classification 
\cite{Komiske_2017,Metodiev_2017,Butter_2018}
and anomaly detection \cite{Farina_2020,Hajer_2020}
in the analysis of particle physics experiments,
aiding observations in astronomy \cite{Fluke_2019}, 
and the study of phases in condensed matter systems 
\cite{Wetzel_2017,Carrasquilla_2017,van_Nieuwenburg_2017,Scheurer_2020}. 
The focus here on the last topic, phases of matter.

Many different machine learning methods have been applied to the exploration of phases
and phase transitions. These include neural networks 
\cite{Wetzel_2017b,Suchsland_2018,Huembeli_2018,Yoshioka_2018},
principal component analysis \cite{PhysRevB.94.195105,Hu_2017},
support vector machines \cite{Ponte_2017}
and diffusion maps \cite{Rodriguez_Nieva_2019,wang2020unsupervised,Long_2020}.
In an interesting recent work, Yang \emph{et al.} \cite{yang2020visualizing} 
suggested the use of t-distributed stochastic neighborhood embedding (tSNE) 
as an unsupervised learning method to obtain a visualization of phase diagrams
and used this method to study a number of one-dimensional quantum spin systems.

The idea of using unsupervised learning methods to reveal phases is 
particularly appealing. Specifically, topological phases
\cite{RevModPhys.88.035005} which can not be characterized by local 
order parameters can be studied without \emph{a priori} input of domain knowledge.
Ref. \cite{wang2020unsupervised,Long_2020,Che_2020,zhang2020machine,narayan2020machine,
kerr2020automatic, yu2020unsupervised,kaming2021unsupervised}
are examples of recent work on identifying topological phases and phase 
transitions using either neural networks or diffusion maps. The Ref.
\cite{Long_2020,Che_2020,zhang2020machine,narayan2020machine,
kerr2020automatic, yu2020unsupervised} focus on the 
Su-Schrieffer-Heeger (SSH) model \cite{PhysRevLett.42.1698,PhysRevB.22.2099}
which is also the subject of this work. 

The SSH model was introduced as a model for polyacetylene and is a simple
extensively studied example of a model for a topological insulator 
\cite{Asb_th_2016, Batra_2020}. The basic model consists of electrons (taken to
be spinless) hopping on a one-dimensional lattice. The nearest-neighbor hopping 
amplitudes are taken to be staggered (see, for example, Fig. 1.1 in
\cite{Asb_th_2016}) so the lattice can be divided into two-site units cells.
Some details of the model are given in Sec. \ref{sec_ssh}.

The SSH model can be extended by introducing longer range interactions. The extended
SSH model considered here allows for the addition of next-next-nearest neighbor
hopping terms \cite{kerr2020automatic,Hsu_2020}. A different type of
modification of the model which has received considerable recent attention
is to allow nonreciprocal intra-cell hoping \cite{Li_2014,Yin_2018}. This leads 
to a non-Hermitian Hamiltonian and the appearance of topological phases
with fractional winding number \cite{Yin_2018}.

In this paper the tSNE algorithm, which has been used to study one-dimensional 
spin systems \cite{yang2020visualizing}, is applied to the SSH model and 
the variants mentioned above. The algorithm is based on dimensionality reduction. 
The model, sampled at a variety of points in its parameter space, is described 
by points distributed in a high-dimensional space and tSNE maps this space to a 
lower-dimensional space where, ideally, there is a clustering of data which can 
be used to identify regions of parameter space which share common features. In 
this work a model which allows calculations to be made for different parameters 
but the same strategy can be applied to data from experiments where measurements 
are made under different experimental conditions.

The implementation of tSNE analysis requires choices to made, for example,
for a distance function for points in the high-dimensional input space and
for the algorithm to identify clusters in the tSNE output. These issues are
discussed in Sec. \ref{sec_res} and in the Appendices. With appropriate 
choices it is found that the unsupervised tSNE analysis can give a correct
visualization of the phase diagram even the most challenging case of the
non-Hermitian extended SSH model where five phases are present.

The tSNE algorithm is introduced in Sec. \ref{sec_tsne} and Sec. \ref{sec_ssh} 
gives a brief outline of the SSH model and some of its extensions. Results are 
presented in Sec. \ref{sec_res} with a summary in Sec. \ref{sec_summ}.

\section{T-DISTRIBUTED STOCHASTIC NEIGHBORHOOD EMBEDDING}
\label{sec_tsne}

T-distributed stochastic neighborhood embedding \cite{JMLR:v9:vandermaaten08a}
(tSNE) is a based
on the notion of dimensionality reduction. The system to be analyzed
is described by a set of points $\left\{ \mathbf{x}\right\} $ in
a space with large dimension which may obscure the presence features
with a much smaller dimensionality. The idea of tSNE is to construct
a mapping of $\left\{ \mathbf{x}\right\} $ to points $\left\{ \mathbf{y}\right\} $
in a space of low dimension (typically 2 or 3) in such a way that
points in the high-dimensional space that share some common feature
will lie close together in the low-dimensional space.

A key element of tSNE, common to other some machine learning algorithms,
for example, diffusion maps, is the distance $D_{ij}(\mathbf{x}_{i},\mathbf{x}_{j})$
between points in the high-dimensional space. The distance $D$ is
commonly taken to be the Euclidean distance between the points but
other metrics can be used and, depending on the application, may be
more effective. The first step of the algorithm is to assign a conditional
probability that point i should have point j as a neighbor
\begin{equation}\label{eq:pji}
P_{j|i}=\frac{e^{-D_{ij}^{2}/2\sigma_{i}^{2}}}{\sum_{_{k\neq i}}e^{-D_{ik}^{2}/2\sigma_{i}^{2}}},
\end{equation}
 for all $i\neq j.$ The $\sigma_{i}$' s are hyperparameters of the
algorithm. Then a probability $P_{ij}$ is defined by
\begin{equation}\label{eq:Pji}
P_{ij}=\frac{P_{i|j}+P_{j|i}}{2N},
\end{equation}
 where $N$ is the number of points in the set $\left\{ \mathbf{x}\right\} $.

A probability distribution $Q_{ij}$ for points $\left\{ \mathbf{y}\right\} $
in the low-dimensional space is also defined. It is taken to be a
Student t-distribution
\begin{equation}\label{eq:Qij}
Q_{ij}=\frac{\left[1+\left\Vert \mathbf{y}_{i}-\mathbf{y}_{j}\right\Vert ^{2}\right]^{-1}}{\sum_{k\neq i}\left[1+\left\Vert \mathbf{y}_{i}-\mathbf{y}_{k}\right\Vert ^{2}\right]},
\end{equation}
 using Euclidean distance. The algorithm then tries to make the probability
distribution Q similar to P by find points $\left\{ \mathbf{y}\right\} $
which minimize the Kullback-Leibler divergence defined by
\begin{equation}
\sum_{ij}P_{ij}\log\frac{P_{ij}}{Q_{ij}}.
\end{equation}
 The tSNE algorithm is included in the machine learning toolkit 
scikit-learn \cite{scikit-learn, tsne}
and that is the implementation that will be used in this work. In
the implementation of tSNE the hyperparameters are fixed implicitly
by requiring that the so-called perplexity $\mathcal{P}$, defined
by
\begin{equation}\label{eq:perplex}
\log_{2}\mathcal{P=}-\sum_{j}P_{j|i}\log_{2}P_{j|i},
\end{equation}
takes a specified value. 

\section{THE SU-SCHRIEFFER-HEEGER MODEL}
\label{sec_ssh}

The SSH model was introduced as a model for polyacetylene
\cite{PhysRevLett.42.1698,PhysRevB.22.2099}. It consists
of electrons hopping on one-dimensional lattice with a two-site unit
cell. The intra-cell and inter-cell couplings are taken to be different.
The Hamiltonian for the basic model with L unit cells is
\begin{equation}
H=\sum_{j=1}^{L}\left[t_{1}a_{j}^{\dagger}b_{j}+t_{2}a_{j+1}^{\dagger}b_{j}\right]+h.c.
\end{equation}

\noindent where $a_{i}^{\dagger}$ and $b_{i}$denote creation and
annihilation operators on the two different sites of the i'th unit
cell. With periodic boundary conditions the Hamiltonian can be written
in momentum space as 
\begin{equation}
H(k)=d_{x}(k)\sigma_{x}+d_{y}(k)\sigma_{y},
\end{equation}

\noindent where the $\sigma$'s are Pauli matrices and the Bloch vectors
are
\begin{eqnarray}\label{eq:sshblh}
d_{x}(k) & = & t_{1}+t_{2}\cos(k),\nonumber \\
d_{y}(k) & = & t_{2}\sin(k).
\end{eqnarray}

\noindent The momenta $k$ take values in {[}$0,2\pi].$ The SSH model
has a nontrivial topological property. If $t_{2}>t_{1}$ the vector
$\mathbf{d}$ = $(d_{x},d_{y})$ will rotate about the origin of the
$d_{x}-d_{y}$ plane as $k$ varies from $0$ to $2\pi$ (see, for
example, Fig. 8 in Ref \cite{Batra_2020}). Note that the 
trajectory of the Bloch vectors
passing through the point $\mathbf{d}$ = 0 is the condition that
the model is gapless Ref \cite{Yin_2018}. This can occur when $t_{1}=\pm t_{2}$
which is the boundary in parameter space between different phases.
When $t_{2}>t_{1}$ the winding number $w$ = 1 where for our discretized
lattice
\begin{equation}
w=\frac{1}{2\pi}\sum_{j}^{L}\varDelta\phi(j),
\end{equation}

\noindent where $\varDelta\phi(j)=\left|\phi(j)-\phi(j-1)\right|$mod
$2\pi$ with $\phi_{j}$ equal to the phase of $d_{x}(k)+id_{y}(k)$
for $k=2\pi j/L$. When $t_{2}<t_{1},$the winding number vanishes.

The SSH model can be extended to include next-to-next-nearest neighbor
inter-cell hoping and the coupling between  nearest neighbor cells
and next-to-next-nearest neighbor cells can be taken to be different.
In this case the Bloch vectors take the form 

\noindent 
\begin{eqnarray}\label{eq:esshblh}
d_{x}(k) & = & t_{1}+t_{2}\cos(k)+t_{3}\cos(2k),\nonumber \\
d_{y}(k) & = & t_{2}\sin(k)+t_{3}\sin(2k).
\end{eqnarray}

\noindent In this case the vector $\mathbf{d}$ will trace out a double
loop in the $d_{x}-d_{y}$ plane and the winding number can take values
0, 1 or 2 (See, for example, \cite{kerr2020automatic,Hsu_2020}).

Another extension of the SSH model is to allow for a non-reciprocal
inter-cell hopping strength. This leads to a non-Hermitian Hamiltonian.
In the non-Hermitian extension considered here the Bloch vectors take
the form \cite{Yin_2018,yu2020unsupervised}

\noindent 
\begin{eqnarray}\label{eq:sshnh}
d_{x}(k) & = & t_{1}+t_{2}\cos(k),\nonumber \\
d_{y}(k) & = & t_{2}\sin(k)-i\delta,
\end{eqnarray}
where $\delta$ is a real parameter characterizing the difference
between left and right intra-cell hopping. The condition for the energy
gap to vanish becomes $t_{1}=\pm t_{2}\pm\delta$. In addition to
phases with winding number 0 and 1, there are regions of parameter
space where $w=1/2$.

The extended SSH model (\ref{eq:esshblh}) can also be modified to include 
non-Hermiticity. The Bloch vectors become \cite{Yin_2018}

\noindent 
\begin{eqnarray}\label{eq:esshnh}
d_{x}(k) & = & t_{1}+t_{2}\cos(k)+t_{3}\cos(2k),\nonumber \\
d_{y}(k) & = & t_{2}\sin(k)+t_{3}\sin(2k)-i\delta.
\end{eqnarray}

\noindent The phase diagram becomes quite complicated with regions
of winding number 1/2 and 3/2 appearing along with winding numbers
0, 1 and 2 that are present when $\delta$ = 0.

\section{Results}
\label{sec_res}

\subsection{Bloch vector input}

In this section results of applying the tSNE reduction to the SSH
models of Section are described. The Bloch vectors calculated at different
points lying in a two-dimensional plane of parameter space and spanning
different phases are used as input into the algorithm. The reduced
space is taken to be two-dimensional. As mentioned in Sec. \ref{sec_tsne} the
distance function in the space of input vectors need not be the Euclidean
distance. Scikit-learn provides a variety of metrics. In this work
the $\mathbb{L}_{p}$ norms with $p=1,2,\infty$ were considered. In
scikit-learn these are called Cityblock, Euclidean and Chebyshev respectively.
Recall that the $\mathbb{L}_{p}$ norm of a vector $\mathbf{v}$ is
\begin{equation}
\mathbb{L}_{p}=\left(\sum_{i}\left|v_{i}\right|^{p}\right)^{\frac{1}{p}}.
\end{equation}
If the number of clusters to which the data are mapped is very small
the choice of metric may not be critical. However, if, for example, the
input data span a large number phases some metrics may be more effective
than others. The example in Appendix \ref{app1} Fig. \ref{fig6} of the 
non-Hermitian extended
SSH model shows that using the Chebyshev distance leads to well separated
clusters while other choices do not. All results presented in this
Section based on Bloch vector input were calculated using the Chebyshev
distance. Note the use of this metric was earlier advocated by 
Che \emph{et al.} \cite{Che_2020}
in their study topological phases using diffusion maps.

\begin{figure}[tbhp]
\centering
\includegraphics[scale=0.60,trim={2cm 1cm 2cm 2cm},clip=false,angle=90]{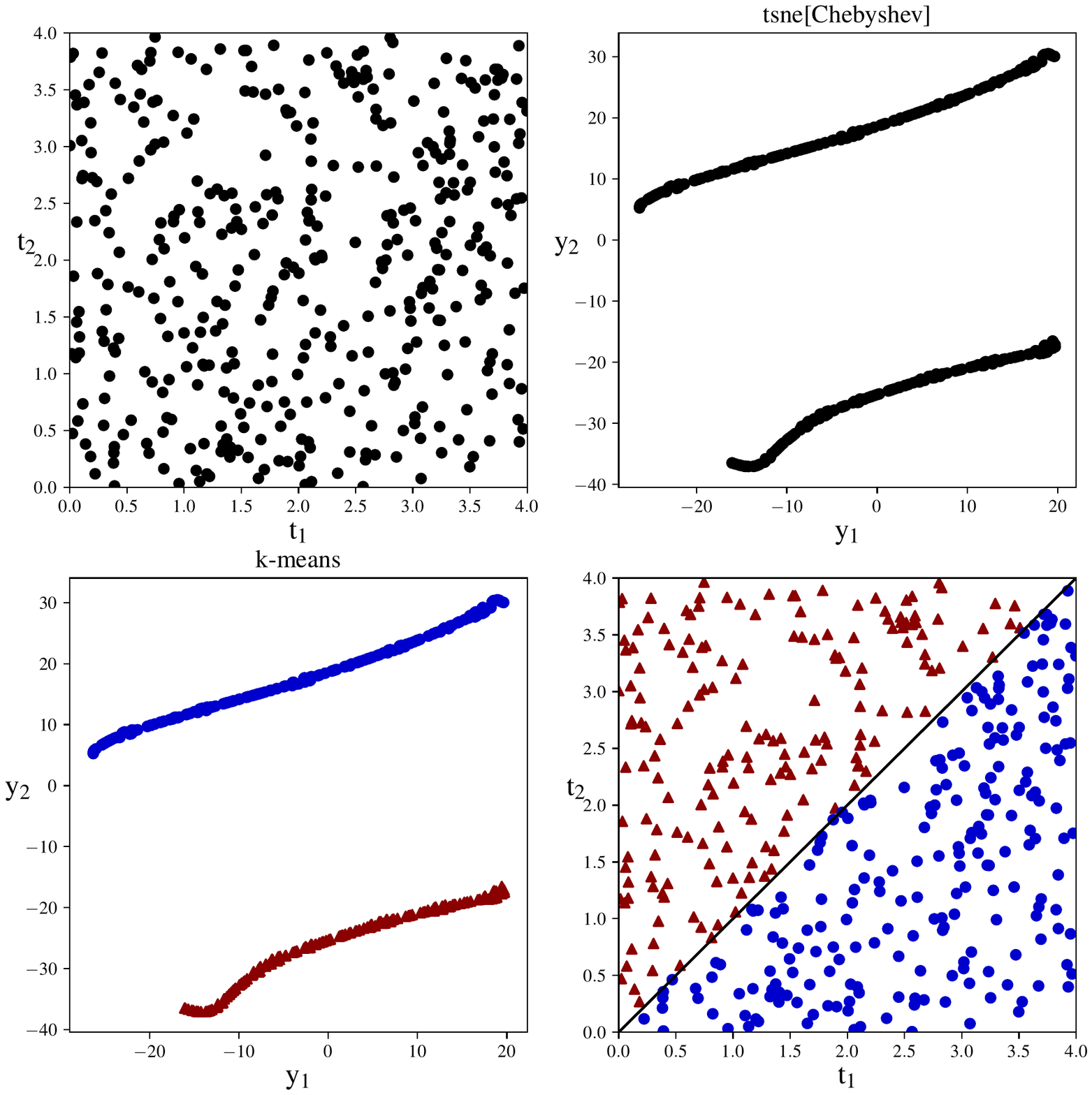}
\caption{Phases of the SSH model exposed by tSNE.Top-left:Points in the $t_1-t_2$
parameter space at which input Bloch vectors are calculated. Top-right:Output of 
tSNE. Bottom-left:Clusters identified by k-means. Bottom-right:Points in
parameter space with color and symbol showing the cluster to which they correspond. The
black line shows the known phase boundary.} 
\label{fig1}

\includegraphics[scale=0.60,trim={1.3cm 1cm 2cm 2cm},clip=true,angle=90]{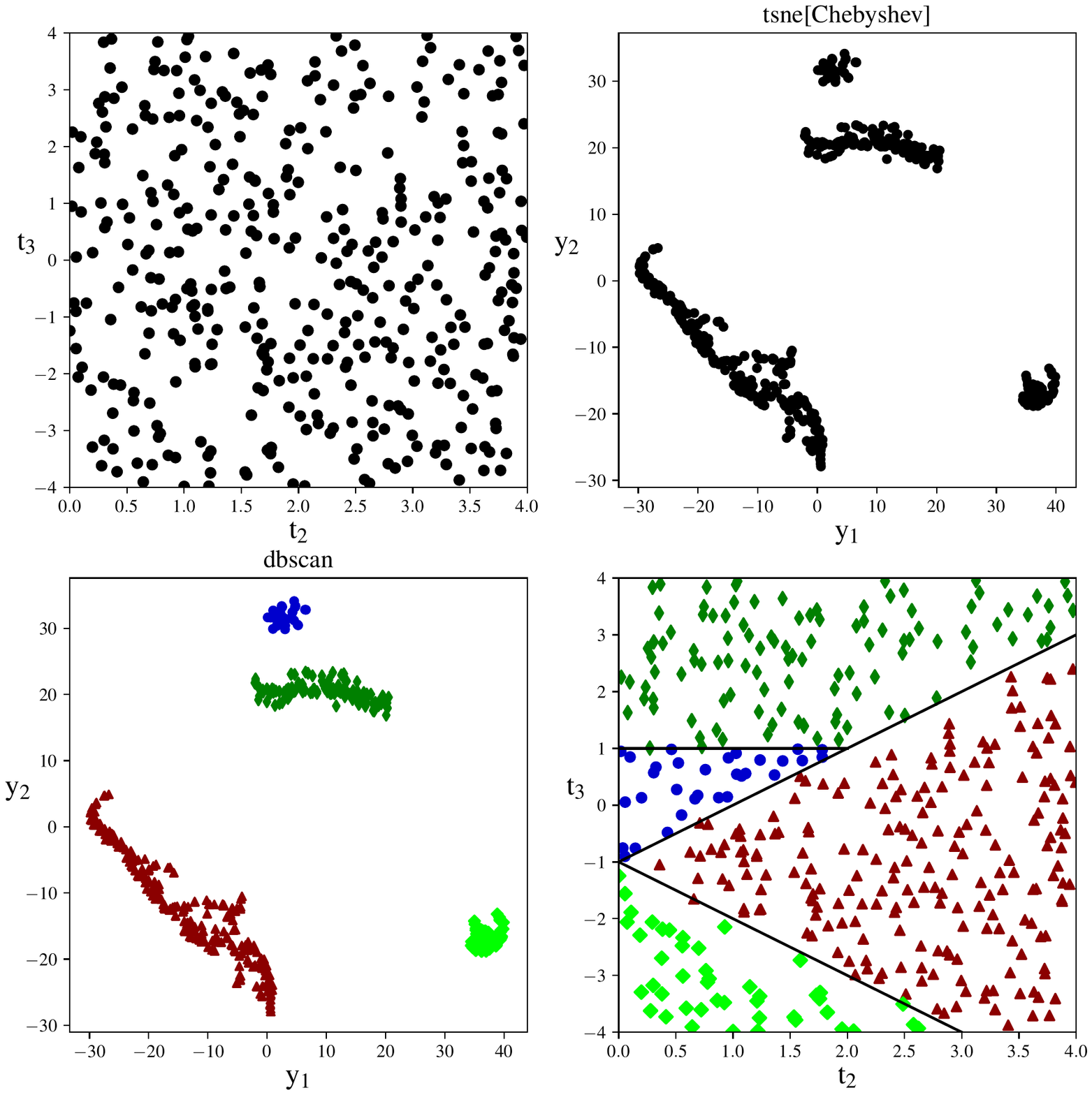}
\caption{Phases of the extended SSH model exposed by tSNE.Top-left:Points in the $t_2-t_3$
parameter space with $t_1$ = 1 at which input Bloch vectors are calculated. Top-right:Output of 
tSNE. Bottom-left:Clusters identified by dbscan. Bottom-right:Points in
parameter space with color and symbol showing the cluster to which they correspond. The
black lines show the known phase boundaries.} 
\label{fig2}
\end{figure}

\begin{figure}[tbh]
\centering
\includegraphics[scale=0.6,trim={2cm 1cm 2cm 2cm},clip=false,angle=90]{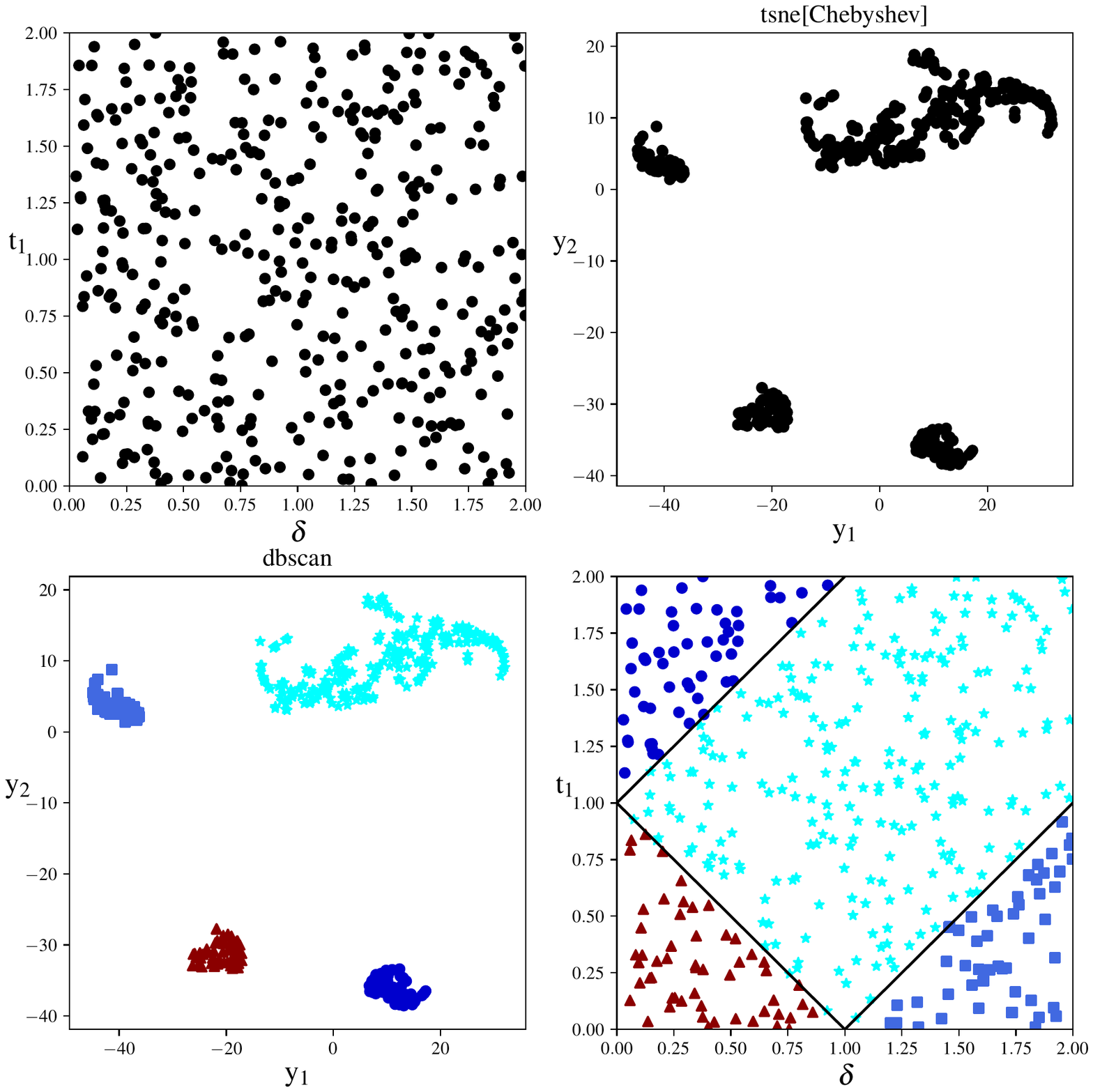}
\caption{Phases of the non-Hermitian SSH model exposed by tSNE.Top-left:Points in the $\delta-t_1$
parameter space with $t_1$ = 1 at which input Bloch vectors are calculated. Top-right:Output of 
tSNE. Bottom-left:Clusters identified by bdscan. Bottom-right:Points in
parameter space with color and symbol showing the cluster to which they correspond. The
black lines show the known phase boundaries.} 
\label{fig3}
\end{figure}

\begin{figure}[tbh]
\centering
\includegraphics[scale=0.6,trim={2cm 1cm 2cm 2cm},clip=false,angle=90]{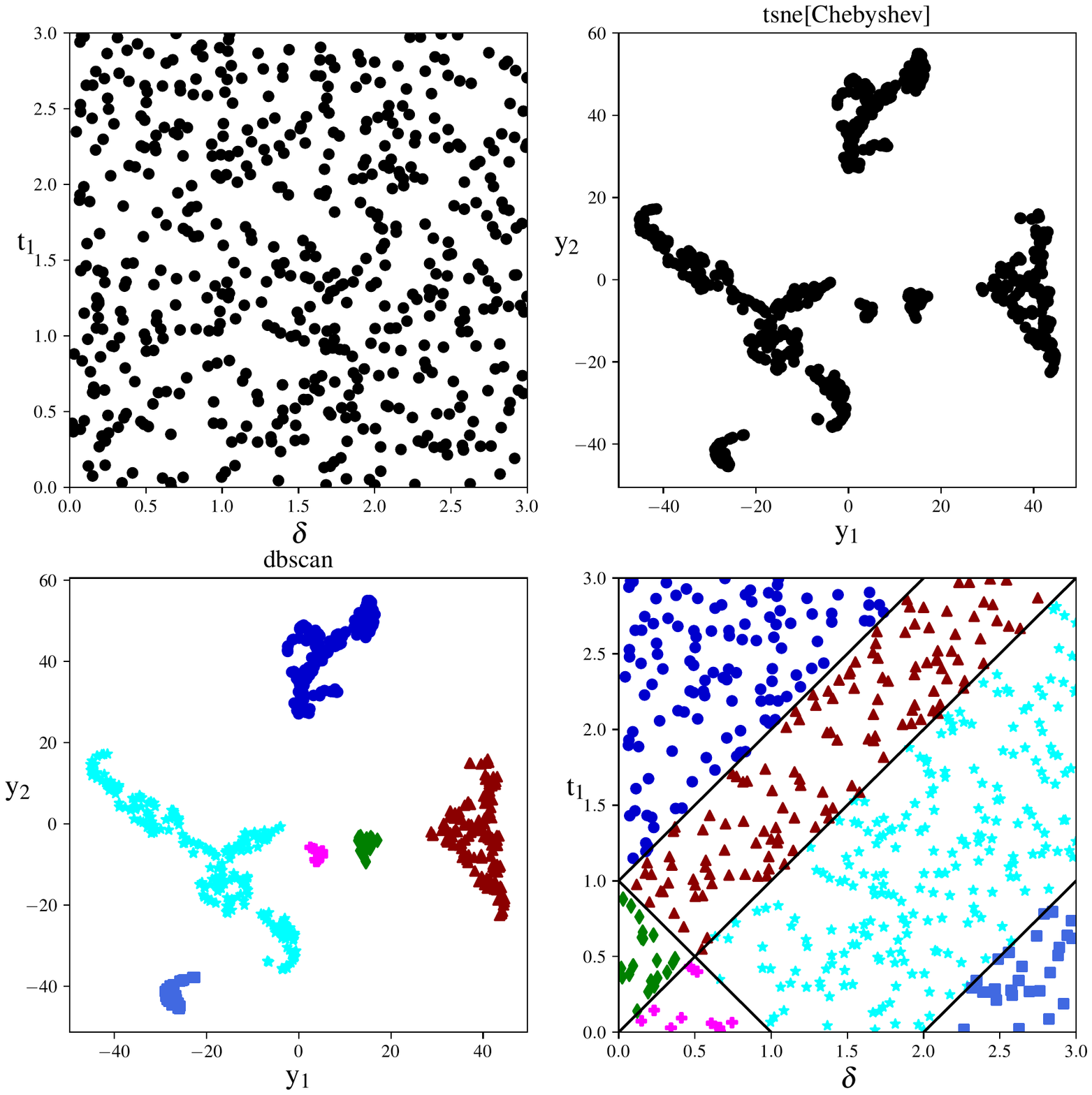}
\caption{Phases of the non-Hermitian extended SSH model exposed by tSNE.Top-left:Points 
in the $\delta-t_1$ parameter space with $t_2,t_3$ = 1 at which input Bloch vectors are calculated. 
Top-right:Output of  tSNE. Bottom-left:Clusters identified by dbscan. Bottom-right:Points in
parameter space with color and symbol showing the cluster to which they correspond. The
black lines show the known phase boundaries.} 
\label{fig4}
\end{figure}

\begin{figure}[tbh]
\centering
\includegraphics[scale=0.6,trim={2cm 1cm 2cm 2cm},clip=false,angle=90]{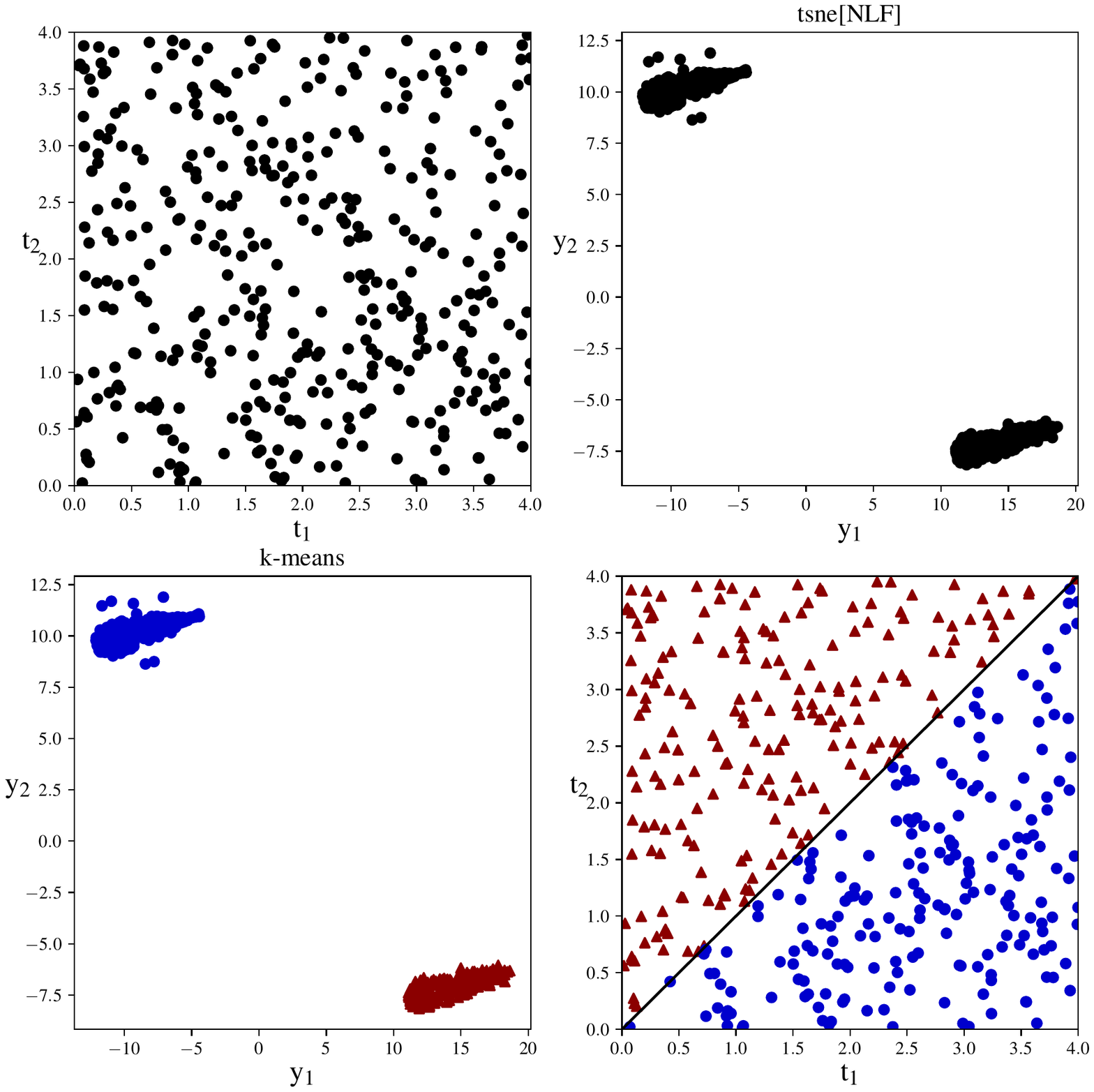}
\caption{Phases of the SSH model exposed by tSNE with wavefunction input.
Top-left:Points in the $t_1-t_2$
parameter space at which input wavefunctions are calculated. Top-right:Output of 
tSNE. Bottom-left:Clusters identified by k-means. Bottom-right:Points in
parameter space with color and symbol showing the cluster to which 
they correspond. The
black line shows the known phase boundary.} 
\label{fig5}
\end{figure}

First consider the basic SSH model Eq. \ref{eq:sshblh}. The model has two parameters.
A sample of 400 Bloch vectors corresponding to a lattice of 80 cells
was constructed using parameters chosen randomly in the range $0\leq t_{1},t_{2}\leq4.$
The selected points are shown in the top-left panel of Fig. \ref{fig1}
The result of tSNE mapping of the Bloch vectors to a two-dimensional space 
using the Chebyshev distance is shown in the top-right panel the figure. The 
default value of the perplexity,  $\mathcal{P}$ = 30, was used.
Two clusters can be identified by inspection or by using a clustering
algorithm such as k-means \cite{dbscan}. The clusters, coded by color and symbol,
are displayed in the bottom left panel. The points in parameter space
are replotted in the bottom-right panel now coded with color and symbol
indicating the cluster to which they were grouped by tSNE. The diagonal
line is the phase boundary between the topological phase $t_{2}>t_{1}$ with
winding number equal to 1 and the band insulator phase $t_{2}<t_{1}$ with
winding number 0. One sees that tSNE makes a clear separation of the
phases in a unsupervised way giving a visualization of the phase diagram.

The extended SSH model, Eq. \ref{eq:esshblh}, has three parameters. We consider 
the model at $t_1 = 1$ and allow the other parameters to vary in the range
$0\leq t_{2} \leq4$ and $-4\leq t_{3} \leq4$. 
The randomly selected points are shown in the 
top-left panel of  Fig. \ref{fig2} and the tSNE output is in the top-right
panel. This model has a more complicated 
phase diagram than the basic SSH model (see Fig. 5 in \cite{kerr2020automatic})
and the choice of clustering algorithm becomes an issue. The results of using
k-means clustering are compared to dbscan \cite{schubert17,dbscan}
in Appendix \ref{app2} Fig. \ref{fig8}. The dbscan clusters,
coded with different colors and symbols, are shown in the bottom-left of panel 
Fig. \ref{fig2}. The corresponding points in the $t_2 - t_3$ plane along with 
the known phase boundaries \cite{kerr2020automatic} are plotted in the 
bottom-right panel. The tSNE analysis gives a clean separation of the 
phase diagram into four regions in agreement with known results. Note that 
this would not be the case if k-means were used for cluster identification.

The phase diagram of the non-Hermitian SSH model , Eq. \ref{eq:sshnh},
is shown in Fig. 3 of Ref. \cite{Yin_2018} for $t_2$ = 1. 
For this work we consider a quadrant of the
parameter space with $0\leq \delta,t_{1} \leq2$. The input Bloch vectors 
were calculated at the 400 points plotted in the top-left panel of  
Fig. \ref{fig3}  and top-right panel shows the tSNE output. Cluster
identification using dbscan is shown in the bottom-left panel. The four
distinct regions in the parameter space are correctly identified (bottom-right panel)
as indicated by the black lines showing the phase boundaries.

The non-Hermitian extended SSH model,  Eq. \ref{eq:sshnh}, presents more of a
challenge. There are five phases with different winding numbers. The phase 
diagram in the  $\delta - t_1$ plane for  $t_2,t_3 = 1$ is given in Fig. 5 of Ref. 
\cite{Yin_2018}. Here 529 points in the range $0\leq \delta,t_{1} \leq3$
are used for the input Bloch vectors. These are shown in the top-left panel
of Fig. \ref{fig4} and the resulting tSNE mapping is in the top-right panel.
A perplexity of 20 was used in this case as it was found to give a better 
separation of the clusters than the default value of 30. The dbscan 
identification of clusters, depicted in different colors and symbols,
is in the lower-left panel. The bottom-right panels shows the points 
in the $\delta - t_1$  plane associated with different clusters. Phase
boundaries are shown by the black lines. It seems quite remarkable that
the combination of tSNE dimensionality reduction and dbscan clustering 
can distinguish all six regions even though some are small and represented by 
only a few points in the input sample.

\subsection{Wavefunction input}

An alternative to using the Bloch vectors to explore the phases is 
to use the eigenvectors of the real-space Hamiltonian. For a lattice
of $L$ cells the Hamiltonian can be expressed as an $2L\times2L$
sparse matrix with nonzero entries along the super- and sub-diagonal
\cite{Asb_th_2016}.
Let $|\psi_{i}>$and $|\psi_{j}>$ denote the lowest positive energy
eigenvectors for two different choices of the model parameters. An
$\mathbb{L}_{p}$ norm of the difference between these vectors can
be used as a distance in an tSNE analysis but this may not be the
best choice. Yang \emph{et al.} \cite{yang2020visualizing} suggest that for 
visualizing quantum phases with tSNE a better measure of distance between 
quantum states would be 
\begin{equation}
-\log\left|<\psi_{i}|\psi_{j}>\right|
\label{eq:nlf}
\end{equation}
which they call the negative logarithmic fidelity (NLF). As an example,
Fig. 7 in Appendix \ref{app1} shows the tSNE reduction 
using a sample of 400 SSH model wavefunctions
calculated on a lattice of 16 cells for different randomly chosen
$t_{1}$ and $t_{2}$ values. Since the model has two phases two clusters
of points in the $y_{1}-y_{2}$ plane are expected. With an $\mathbb{L}_{p}$ 
norm one sees two clusters of outliers in addition to the main clusters
which makes the association of clusters with phases problematic. With
NLF there is a clear separation into reasonably compact clusters.
The complete analysis with wavefunction input using NLF is shown in
Fig. \ref{fig5}. The two phases are clearly distinguished just as in 
Fig. \ref{fig1} where Bloch vectors were used.

\section{Summary and discussion}
\label{sec_summ}

Machine learning offers many ways of exploring phase transitions in 
condensed matter systems. Particularly interesting are unsupervised 
methods which require no \emph{a priori} knowledge such a as choice of an
order parameter. Such methods are especially suited to the study of
topological phase transitions where there is no local order parameter.

In this work it is shown that t-distributed Stochastic Neighborhood Embedding
it is can be used to learn the phase boundaries of the Su-Schrieffer-Heeger
model and some of its extensions. Input into the analysis consists of Bloch
vectors constructed at different points of the model parameter space.
The tSNE algorithm was used to map the Bloch vectors to clusters in a reduced space
(two-dimensional in this work) corresponding to different regions of 
the phase diagram. This allows a visualization of the phase diagram 
as shown in Figs. \ref{fig1} to \ref{fig4}. Wavefunction input can
also be used in the analysis as shown in Fig. \ref{fig5}.

tSNE does not work automatically. The choice of distance function used
to construct the probability distribution of the input can affect the results. 
For Bloch vector input, the Chebyshev distance was preferred since it worked 
well even in the difficult case of the non-Hermitian extended SSH model.
With wavefunction input, negative logarithmic fidelity  \cite{yang2020visualizing}
was found to useful whereas $\mathbb{L}_{p}$ norms did not produce usable
even in the simple case with only two phases. 
As well, the perplexity may require some adjustment to get a good separation 
of the clusters in the reduced space. Since tSNE is stochastic, the 
output varies from run to run so making adjustments to the algorithm 
requires some care.

Although tSNE can be used to identify regions in different phases, it can
not find the properties, such as the winding number, associated to different
regions. This true also for other machine learning methods, for example, 
principal component analysis or diffusion maps, used to expose phase 
boundaries. Nonetheless, since these learning algorithms are unsupervised,
they can provide useful information about phase transitions without 
input of domain knowledge.

\acknowledgments
TRIUMF receives federal funding via a contribution agreement 
with the National Research Council of Canada.

\appendix
\section{Choice of metric}
\label{app1}

\begin{figure}[tbh]
\centering
\includegraphics[scale=0.6,trim={1cm 1cm 2cm 2cm},clip=true,angle=90]{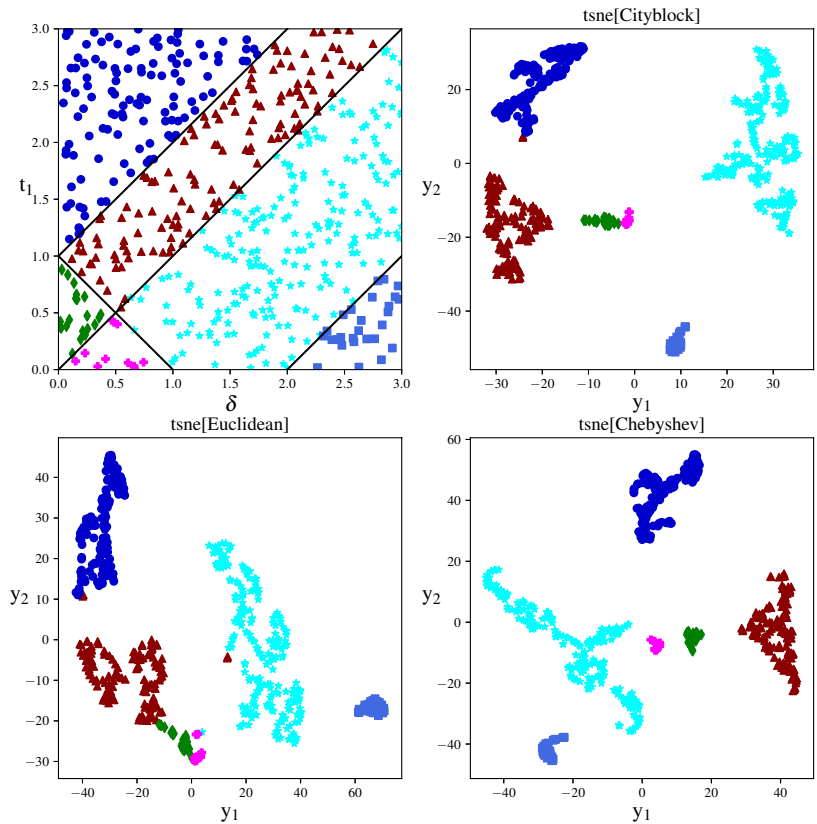}
\caption{tSNE mapping of the non-Hermitian extended SSH model using different
distance functions.Top-left:Points in the $\delta-t_1$ parameter space with $t_2,t_3$ = 1 
at which input Bloch vectors are calculated. The other panels show tSNE output using the 
distance function indicated in the panel title in Eq. \ref{eq:pji}.} 
\label{fig6}
\end{figure}

\begin{figure}[tbhp]
\centering
\includegraphics[scale=0.6,trim={2cm 1cm 2cm 2cm},clip=false,angle=90]{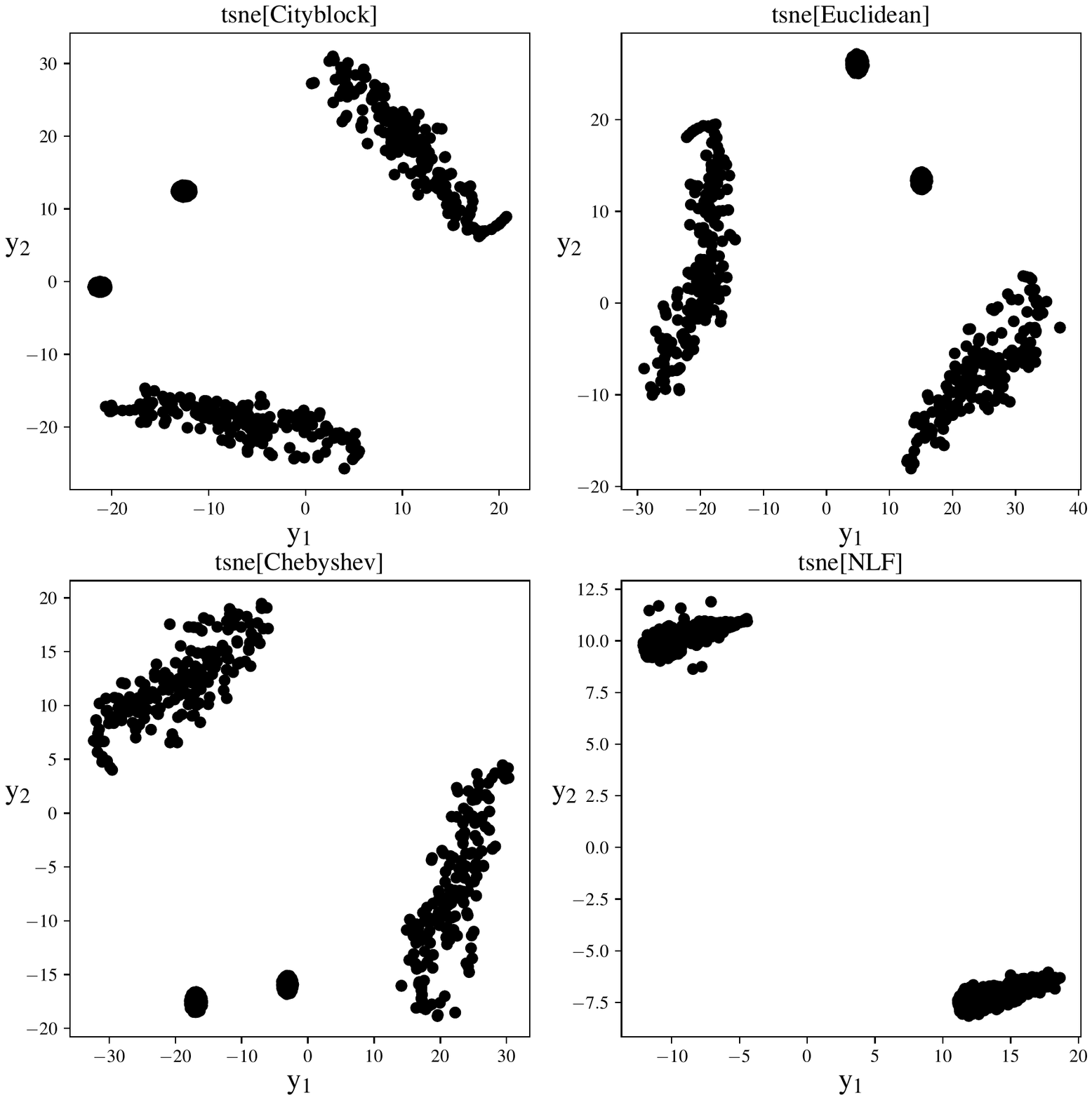}
\caption{tSNE mapping of the SSH model with wavefunction input using the 
distance function indicated in the panel title in Eq. \ref{eq:pji}.}
\label{fig7}

\includegraphics[scale=0.6,trim={2cm 1cm 2cm 2cm},clip=true,angle=90]{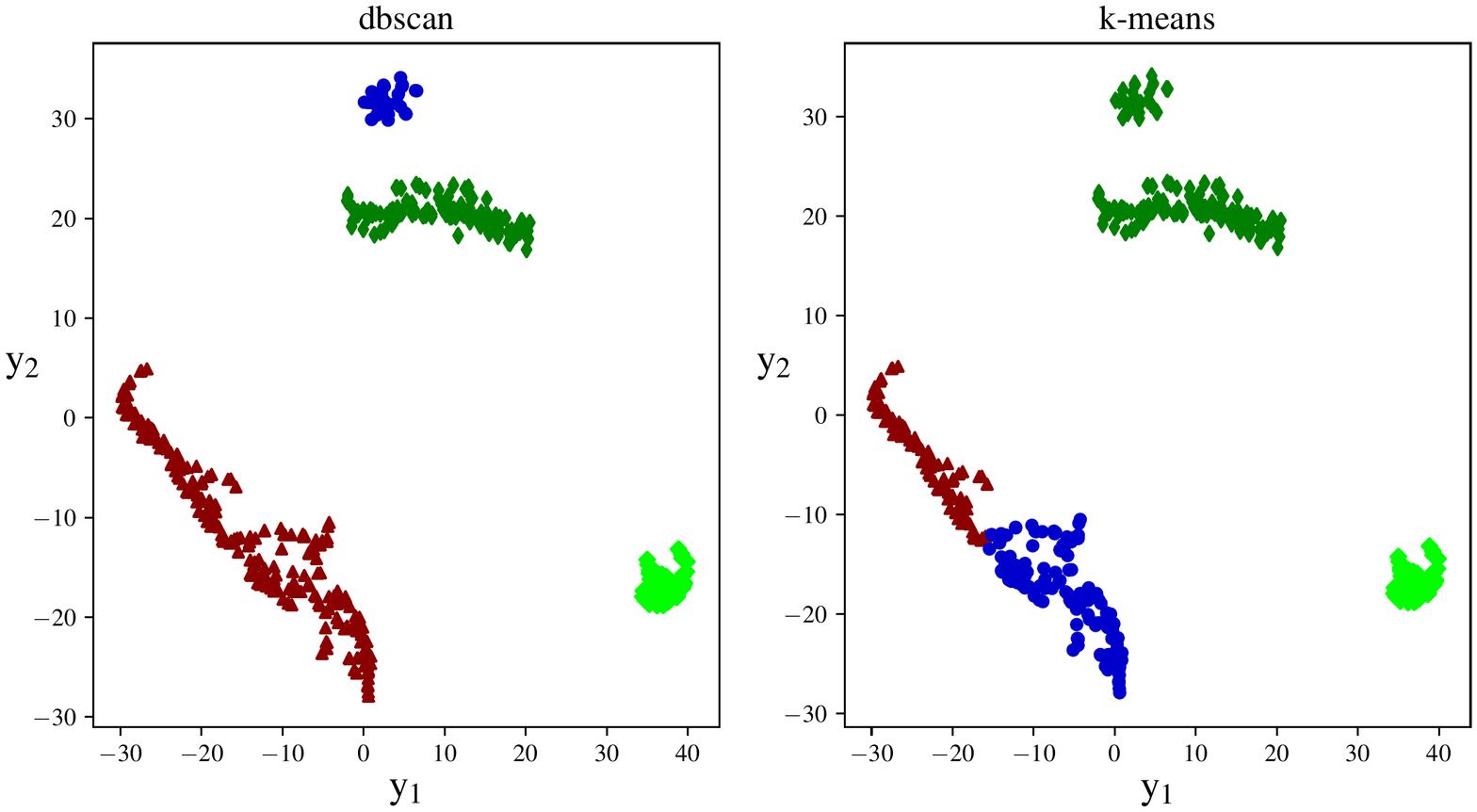}
\caption{Comparison of dbscan and k-means clsutering of tSNE output for the
extended SSH model shown in the top-right panel of Fig. \ref{fig2}.} 
\label{fig8}
\end{figure}

The choice of distance function used in constructing the probability 
distribution Eq. \ref{eq:pji} in the input space can affect the mapping
in the reduced space. As an example the case of the non-Hermitian extended 
SSH model is presented here. Bloch vectors at 529 points selected in the 
range $0\leq \delta,t_{1} \leq3$ with  $t_2,t_3 = 1$ were used as input.
These are plotted in the top-left panel of Fig. \ref{fig6}. The black lines show
phase boundaries. The other panels show the tSNE mapping to two-dimensional 
space where the distance function used is indicated in the panel title.
Cityblock is the scikit-learn name for the  $\mathbb{L}_{1}$ norm. Only the 
Chebyshev distance leads to a usable clustering corresponding to the 
different regions of the phase diagram. Cityblock clusters the points 
correctly but does provide any separation between points mapped from
the two small regions indicated by diamonds and +'s. In this example 
the input data are labeled to illustrate the effect of different 
distance functions. In an analysis where the input is not labeled 
separation of clusters from different regions is critical in order
to identify phases correctly.

Scikit-learn provides a variety of distance functions but sometimes
a custom distance function may have to be crafted in order to get
good results. Fig. \ref{fig7} shows the tSNE mapping of the SSH model 
for wavefunction input using different distance functions as indicated
in the panel titles. This model has two phases so in the reduced
space one would expect to see two clusters of points if tSNE is 
identifying the phases correctly. This is not case using the 
Cityblock, Euclidean and Chebyshev distances. A binary classification
of the points is problematic. However, using the custom distance function, 
negative logarithmic fidelity \cite{yang2020visualizing}, Eq. \ref{eq:nlf},
gives two reasonably compact well-separated clusters.

\section{k-means versus dbscan}
\label{app2}

After mapping of the input data to a reduced space one would like to
identify clusters of points sharing some common features. This could be
done by inspection but better would be to have an algorithmic  
procedure which more objective. However, this can lead to a problem
if the choice of clustering algorithmic is not appropriate. An example 
is shown in Fig. \ref{fig8}.

The two panels panels show the clusters, indicated 
by different symbols and colors, returned by the dbscan and k-means
clustering algorithms \cite{dbscan} for the tSNE output of the extended 
SSH model (top-right panel of Fig. \ref{fig2}). The tSNE output has 
well separated clusters but they are not particularly compact. 
The k-means clustering algorithm, which calculates distances from a set of 
centroids, fails in this case where there is a cluster that is quite extended. 
On the other hand, dbscan, which analyzes the data by dividing it into
subgroups based on the density of points, gives the correct identification 
of clusters associated with different regions of the phase diagram as shown 
in Fig. \ref{fig2}.

\bibliographystyle{h-physrev4}
\bibliography{sshbib}

\end{document}